\documentclass[twocolumn,preprintnumbers,amsmath,amssymb]{revtex4}

\usepackage[english]{babel}
\pdfoutput=1
\usepackage{graphicx,amsfonts}
\usepackage{amsmath}
\begin{document}

\title{Evolutionary Dynamics of Populations with Conflicting Interactions:\\
Classification and Analytical Treatment Considering Asymmetry and Power}

\author{Dirk Helbing$^{1,2,3}$}\email{dhelbing@ethz.ch}
\author{Anders Johansson$^1$}
\affiliation{$^1$ ETH Zurich, CLU, Clausiusstr. 50, 8092 Zurich, Switzerland}
\affiliation{$^2$ Santa Fe Institute, 1399 Hyde Park Road, Santa Fe, NM 87501, USA}
\affiliation{$^3$ Collegium Budapest---Institute for Advanced Study, Szenth\'{a}roms\'{a}g u. 2, 1014 Budapest, Hungary}

\date{\today}
\begin{abstract}Evolutionary game theory has been successfully used to investigate the dynamics of systems, in which many entities have competitive interactions. From a physics point of view, it is interesting to study conditions under which a coordination or cooperation of interacting entities will occur, be it spins, particles, bacteria, animals, or humans.  Here, we analyze the case, where the entities are heterogeneous, particularly the case of two populations with conflicting interactions and two possible states. For such systems, explicit mathematical formulas will be determined for the stationary solutions and the associated eigenvalues, which determine their stability. In this way, four different types of system dynamics can be classified, and the various kinds of phase transitions between them will be discussed. While these results are interesting from a physics point of view, they are also relevant for social, economic, and biological systems, as they allow one to understand conditions for (1) the breakdown of cooperation, (2) the coexistence of different behaviors (``subcultures''), (3) the evolution of commonly shared behaviors (``norms''), and (4) the occurence of polarization or conflict. We point out that norms have a similar function in social systems that forces have in physics.
\end{abstract}


\maketitle


\medskip

\section{Introduction}

Game theory is a theory of interactions, which goes back to von Neumann \cite{Neumann}, one of the superminds of quantum mechanics. It is is based on mathematical analyses \cite{gamedyn0,Weibull,Cressman,NowBook} and methods from statistical physics and the theory of complex systems \cite{HelGame,minority,Roca,cyclic,Claus}, while applications range from biology \cite{gamedyn0,NowBook} over sociology \cite{Axelrod1,gamedyn3,Skyrms,Boyd,Bounds} to economics \cite{Neumann,Bounds,Binmore,Henrich}. Physicists have been particularly interested in {\it evolutionary} game theory \cite{gamedyn0,Weibull,Cressman,gamedyn3,exploratory}, which focuses on the dynamics resulting from the interactions among a large number of entities. These could, for example, be spins, particles, bacteria, animals, or human beings. For such systems, one can calculate the statistical distribution of states in which the entities can be. These states reflect, for example, the location in space \cite{NJP,EPL} and/or whether a spin is oriented ``up'' or ``down'' \cite{RandomReplicators,Preprint},
while in non-physical systems, the states represent decisions, behaviors, or strategies. In such a way, one can study problems ranging from the spontaneous magnetization in spin glasses \cite{RandomReplicators,Preprint} up to the emergence of behavioral conventions \cite{HelGame,Mueller,Peyton}. Further application areas are nucleation processes \cite{Schweitz1,Schweitz2}, the theory of evolution \cite{Eigen,Fisher,Ebeling,gamedyn0}, predator-prey systems \cite{Hofbauer,Montroll} and the stability of ecosystems \cite{Montroll,MayBook,Diversity,ReplicatorEcosyst3}.
Physicists have also been interested in the effects of spatial interactions \cite{Space,Space2,PhysLife}
or network interactions \cite{Network2,networkSW,coevol,Zhong,network,Ohtsuki,Communities,Szolnoki}, of mobility \cite{Frey,NJP,HelPlat,HelACS,HelEPJB,EPL,HelPNAS} or perturbations \cite{EPL,HelPNAS,Perc,Noise2,Noise3}.
\par
Recently, particular attention has been paid to the emergence of cooperation in dilemma situations \cite{NowBook,Five}, which are reflected by a number of different games charactarized by different types of interactions \cite{Weibull}: In the {\it stag hunt game (SH)}, cooperation is risky, in the {\it snowdrift game (SD)}, free-riding (``defection'') is tempting, while both problems occur in the {\it prisoner's dilemma (PD)} \cite{PhysLife}. Details will be discussed in Sec. \ref{CLASSI}.
Most of the related studies have assumed homogeneous populations so far (where every entity has the same kind of interactions). Here, we will study the heterogeneous case with multiple interacting populations. Compared to previous contributions for multiple populations \cite{Mueller,SelfInt,Weibull,Arga,Kana2}, we will focus on populations with conflicting interests and different power. Furthermore, we will classify the possible dynamical outcomes, and discuss the phase transitions when model parameters cross certain critical thresholds (``tipping points'').
\par
Our paper is structured as follows: Section \ref{GADY} introduces the game-dynamical replicator equations for multiple interacting populations. Afterwards, Sec. \ref{CONF} specifies the payoff matrices representing conflicting interactions. While doing so, we will take into account the (potentially different) power of populations. Then, Sec. \ref{STAT} derives the stationary solutions of the evolutionary equations  and the associated eigenvalues, which determine the instability properties of the stationary solutions. This is the basis of our classification. Section \ref{MAIN} collects and discusses the main results regarding the dynamics of the system and possible phase transitions when model parameters are changing. It also offers an interpretation of the formal theory. Finally, Sec. \ref{SUM} presents a summary and outlook.

\section{Game-Dynamical Replicator Equations for Interacting Populations}\label{GADY}

In the following, we will formulate game-dynamical equations for multi-population
interactions \cite{Mueller,SelfInt,Weibull,Arga,Kana2}. For this, we will distinguish
different (sub-)populations $a, b, c \in\{1,\dots,\mathcal{A}\}$
and various states (behaviors, strategies) $i, j, k \in
\{1,\dots,I\}$. If an entity of population $a$ characterized by state
$i$ interacts with an entity of population $b$ characterized by state
$j$, the outcome (``success'') of the interaction is quantified by the
``payoff'' $A_{ij}^{ab}$. Now, let $f_a\ge 0$ with
$\sum_a f_a =1$ be the fraction of entities belonging to
population $a$ and $p_i^a(t)\ge 0$ with $\sum_i p_i^a(t) =1$ the
proportion of entities in population $a$ characterized by state $i$ at
time $t$. We will assume that entities take over (copy, imitate) states that
are more successful in their population in accordance with the
proportional imitation rule \cite{Mueller,Schlag}. Moreover, when the
interaction frequency with entities of population $b$ characterized by
state $j$ is $f_b p_j^b$ (i.e. proportional to the relative size or ``power''
$f_b$ of that population and the relative frequency $p_j^b$ of
state $j$ in it), we find the following set of coupled
game-dynamical equations \cite{Mueller}:
\begin{equation}
\frac{dp_i^a(t)}{dt} = p_i^a(t) \big[ E_i^a(t) - A_a(t) \big] \, .
\label{repli}
\end{equation}
Herein, the ``expected success''
\begin{equation}
E_i^a(t) = \sum_{b=1}^{\mathcal{A}} \sum_{j=1}^I A_{ij}^{ab} f_b p_j^b(t)
\label{repli3}
\end{equation}
of entities belonging to population $a$ characterized by state $i$ is
obtained by summing up the payoffs $A_{ij}^{ab}$ over all possible
states $j$ of interaction partners and populations $b$,
weighting the payoffs with the respective occurrence frequencies
$f_b p_j^b(t)$. (Note that $\sum_b\sum_j f_b p_j^b(t) = 1$.) The
quantity
\begin{equation}
A_a(t) = \sum_{k=1}^I  p_k^a(t) E_k^a(t) \label{repli4}
\end{equation}
is the average success in population $a$ and
\begin{equation}
\langle A \rangle = \sum_{a=1}^{\mathcal{A}} f_a A_a(t)
\end{equation}
the average success in all populations. The above game-dynamical
equations assume that population sizes (and the population an entity
belongs to) do not change.
\par\begin{figure}[htbp]
\includegraphics[width=8.8cm]{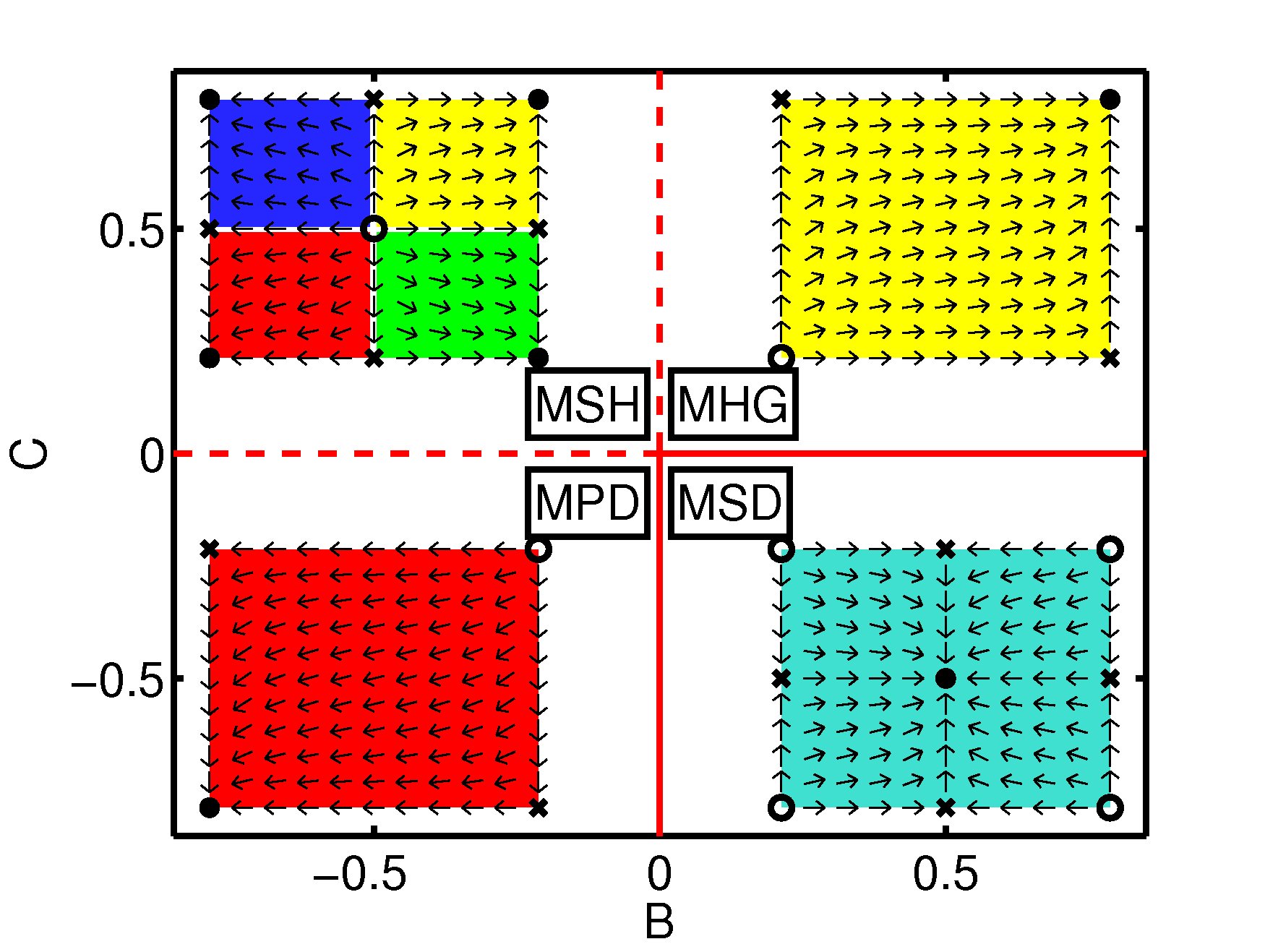}
\caption{(Color Online)
Illustration of the outcomes of symmetrical $2\times2$ games as a function of the payoff-dependent parameters $b_a=B$ and $c_a=C$, if $f=0.8$ (i.e. 80\% of individuals belong to population 1)
and if the entities interact within their own population, but different populations do not have
any interactions between each other ($B_a=0=C_a$) \cite{Weibull,Preprint}.
$p=p_1^1$ is the fraction of entities of population 1 in state 1 and $q=p_2^2$ the fraction of entities of population 2 in state 2. The vector fields show $(dp/dt,dq/dt)$, i.e. the direction and size of the expected change of the distribution $(p,q)$ of states with time $t$. Sample trajectories illustrate some representative flow lines $(p(t),q(t))$ as time $t$ passes. The flow lines move away from unstable stationary points (empty circles). Saddle points (crosses) are attractive in one direction, but repulsive in another. Stable stationary points (black circles) attract the flow lines from all directions. Each color (grey shade) represents one basin of attraction. It subsumes all initial conditions $(p(0),q(0))$ leading to the same stationary point [yellow = (1,1), green = (1,0), blue = (0,1), red = (0,0), turquoise = $(p_0,p_0)$ with $p_0 = |B|/(|B|+|C|)$]. Solid red lines indicate the thresholds at which continuous (``second-order'') phase transitions take place, i.e. at which the system behavior changes qualitatively (characterized by the appearance or disappearance of stationary points), while the {\it stable} stationary points change continuously when the parameters are varied. Dashed lines indicate an abrupt change of a stable stationary point, i.e. a discontinous (``first-order'') phase transition. For multi-population prisoner's dilemmas (MPD), we have $B<0$ and $C<0$, and the final outcome is $(p,q)=(0,0)$. For multi-population snowdrift games (MSD),  we have $B>0$ and $C<0$, and the stable stationary solution corresponds to a coexistence of a fraction $p_0 = |B|/(|B|+|C|)$ of entities in one state and a fraction $1-p_0$ of entities in the other. For multi-population harmony games (MHG), we have $B>0$ and $C>0$, and the eventually resulting outcome is (1,1). Finally, for multi-population stag hunt games (MSH), we have $B<0$ and $C>0$, and there is a bistable situation, i.e. it depends on the initial fraction of entities in a state, whether everybody ends up in this state or in the other one  \cite{Preprint}.}
\label{FIG1}
\end{figure}
Comparing the above game-dynamical equations with the usual
replicator equation for the one-population case, we have additional
terms involving payoffs $A_{ij}^{ab}$ from interactions with {\it
different} populations $b\ne a$. They lead to a mutual coupling of
the replicator equations (\ref{repli}). {\it Asymmetrical} games with
different payoff matrices of the interacting entities, or games
between entities with different sets of states (strategy sets) are examples for
the need to distinguish between {\it different} populations. Within
the framework of game-dynamical equations they can be treated as
{\it bimatrix} games \cite{Weibull,gamedyn0,Cressman}. These,
however, do not consider interactions among entities belonging to
the {\it same} population (``self-interactions''), which are
reflected by the payoff matrices $A_{ij}^{aa}$. The above
multi-population replicator equations include interactions both
within the {\it same} population and between {\it different}
populations. The significantly different dynamics and outcomes when
interactions {\it between} two populations are neglected or when
{\it self-interactions} are neglected become obvious when
Figs. 1 and 2 are compared with Fig. 3.
\par
\begin{figure}[htbp]
\includegraphics[width=8.8cm]{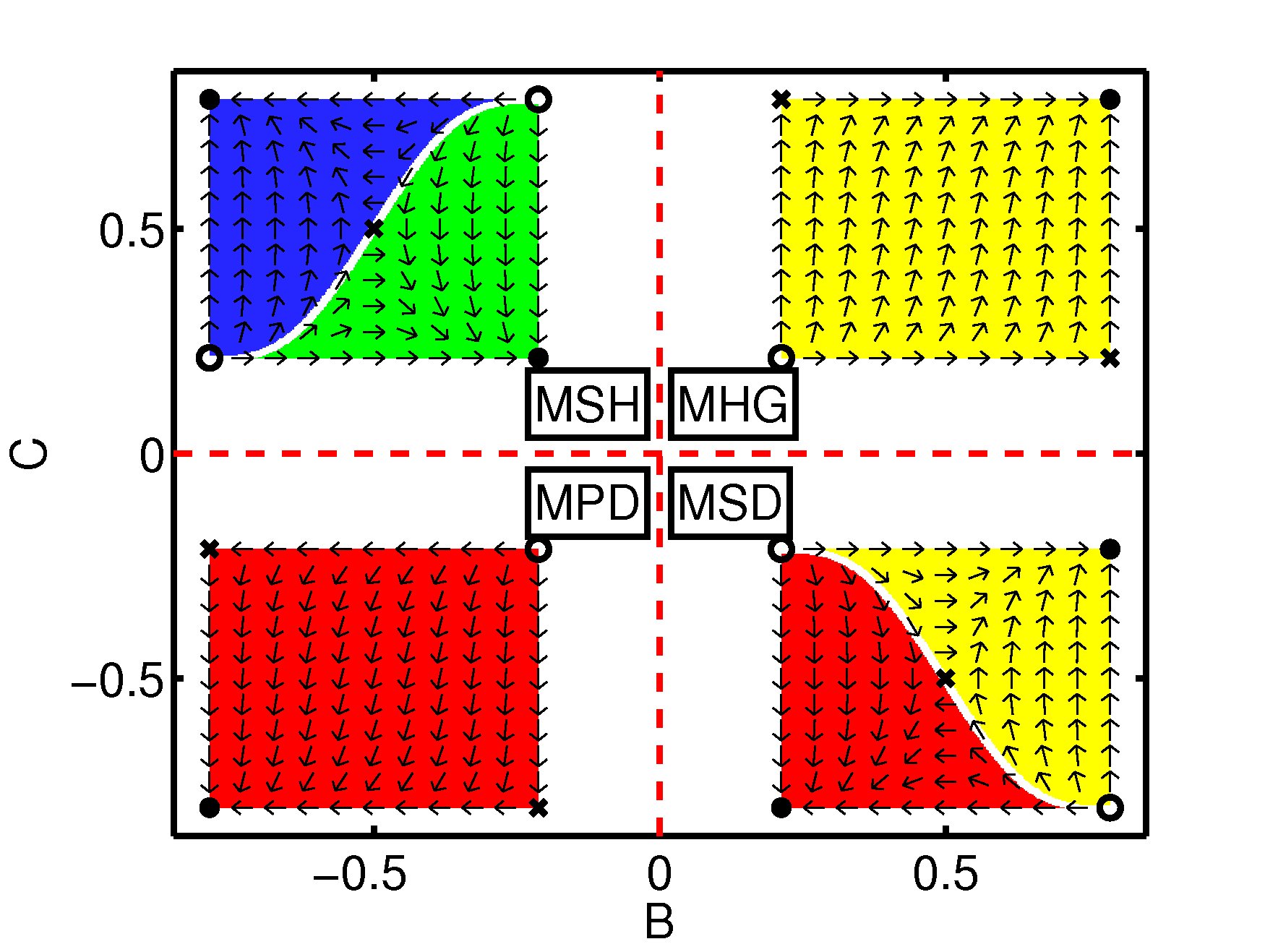}
\caption{(Color Online) Illustration of the outcomes as a function of the payoff-dependent parameters $B_a=B$ and $C_a=C$, if $f=0.8$ (i.e. 80\% of the entities belong to population 1) and if the entities do not interact within their own population ($b_a=0=c_a$), while entities belonging to different populations have interactions with each other \cite{Weibull}. Small arrows illustrate again the vector field $(dp/dt,dq/dt)$ as a function of $p=p_1^1$ and $q=p_2^2$. Black circles represent stable fix points, empty circles
stand for unstable fix points, and crosses represent saddle points. The basins of attraction of different
stable fix points are represented in different grey shades (colors) [yellow = (1,1), green = (1,0), blue = (0,1), red = (0,0)]. Solid red lines indicate the thresholds at which continuous phase transitions
take place, dashed lines indicate discontinous phase transitions. For multi-population prisoner's dilemmas (MPD), we have $B<0$ and $C<0$, for multi-population snowdrift games (MSD),  we have $B>0$ and $C<0$, for multi-population harmony games (MHG), we have $B>0$ and $C>0$, and for the multi-population stag hunt game (MSH), we have $B<0$ and $C>0$.}
\label{FIG2}
\end{figure}
For reasons of simplicity and analytical tractability, we will now
focus on the case of {\it two} populations ($\mathcal{A}=2$) with two
states each ($I=2$). This allows one to reduce the number of
variables by means of the normalization conditions $f_1 = 1-f_2$,
$p_2^1(t) = 1 - p_1^1(t)$ and $p_1^2(t) = 1-p_2^2(t)$. Furthermore,
we find
\begin{eqnarray}
E_1^1(t) - A_1(t) &=& E_1^1(t) - p_1^1(t) E_1^1(t) - [1-p_1^1(t)]
E_2^1(t) \nonumber \\
&=& [1-p_1^1(t)] [E_1^1(t)-E_2^1(t)] \, . \label{inse3}
\end{eqnarray}
When evaluating the expected success $E_i^a(t)$, we will write the
payoff matrices $A_{ij}^{ab}$ for population $a=1$ as
\begin{equation}
(A_{ij}^{11}) = \left(
\begin{array}{cc}
r_1 & s_1 \\
t_1 & p_1
\end{array}\right)
\quad \mbox{and} \quad
 (A_{ij}^{12}) = \left(
\begin{array}{cc}
R_1 & S_1 \\
T_1 & P_1
\end{array}\right) \, .
\end{equation}

\subsection{Specification of Conflicting Interactions}\label{CONF}

To reflect conflicting interactions, the
payoffs in population $a=2$ are assumed to be inverted
(``mirrored''), i.e. state 2 plays the
role in population 2 that state 1 plays in population 1:
\begin{equation}
(A_{ij}^{21}) = \left(
\begin{array}{cc}
P_2 & T_2 \\
S_2 & R_2
\end{array}\right)
\quad \mbox{and} \quad
 (A_{ij}^{22}) = \left(
\begin{array}{cc}
p_2 & t_2 \\
s_2 & r_2
\end{array}\right) \, .
\end{equation}
With the abbreviations $p(t)=p_1^1(t)$ and $q(t)=p_2^2(t)$,
this leads to
\begin{eqnarray}
E_1^1(t) &=& r_1 f p(t) + s_1 f [1 - p(t)] \nonumber \\
&+& R_1 (1-f) [1-q(t)] + S_1 (1-f) q(t)
\label{inse1}
\end{eqnarray}
and
\begin{eqnarray}
E_2^1(t) &=& t_1 f p(t) + p_1 f [1 - p(t)] \nonumber \\
&+& T_1 (1-f) [1-q(t)] + P (1-f) q(t) \, .
\label{inse2}
\end{eqnarray}
The parameter $f = f_1$ represents the (relative) power of population 1, and $(1-f) =f_2$ the power of
population 2. Inserting Eqs. (\ref{inse1}) and (\ref{inse2}) into Eqs. (\ref{inse3}) and (\ref{repli}),
the game-dynamical equation
for population 1 becomes
\begin{equation}
\frac{dp(t)}{dt} = \underbrace{p(t)[1-p(t)]}_{\rm saturation\
factors} \underbrace{F\big(p(t),q(t)\big)}_{\rm growth\ factor}
\label{sat1}
\end{equation}
with $F(p,q)=E_1^1 - E_2^1$. Explicitly, we have
\begin{equation}
F(p,q) = b_1f + (c_1 - b_1) f p + C_1(1-f) +
(B_1-C_1)(1-f)q \, , \label{into1}
\end{equation}
where
\begin{equation}
b_1 = s_1-p_1, \mbox{ \ }  c_1 = r_1-t_1,  \mbox{ \ }  B_1 = S_1-P_1,  \mbox{ \ }
C_1=R_1-T_1 \, .
\end{equation}
The supplementary  equation for population 2 reads
\begin{equation}
\frac{dq(t)}{dt} = \underbrace{q(t)[1-q(t)]}_{\rm saturation\
factors} \underbrace{G\big(p(t),q(t)\big)}_{\rm growth\ factor}
\label{sat2}
\end{equation}
with
\begin{equation}
G(p,q) = b_2(1-f) + (c_2 - b_2) (1-f) q + C_2f + (B_2-C_2)fp \, .
\label{into2}
\end{equation}
It is obtained by exchanging $p$ and $q$, $f$ and $1-f$, and indices
$1$ and $2$. The first factors may be interpreted as saturation
factors, as they limit the proportions $p$ and $q$ to the admissible
range from 0 to 1. The factors $F(p,q)$ and $G(p,q)$ can be
interpreted as growth factors, if greater than zero (or as decay
factors, if smaller than zero). Note that the above two-population
game-dynamical equations are general enough to capture all possible
$2\times2$ games and even situations when entities of different
populations play different {\it kinds} of games ({\it ``asymmetrical''} case).
\begin{figure*}[htbp]
\includegraphics[width=18cm]{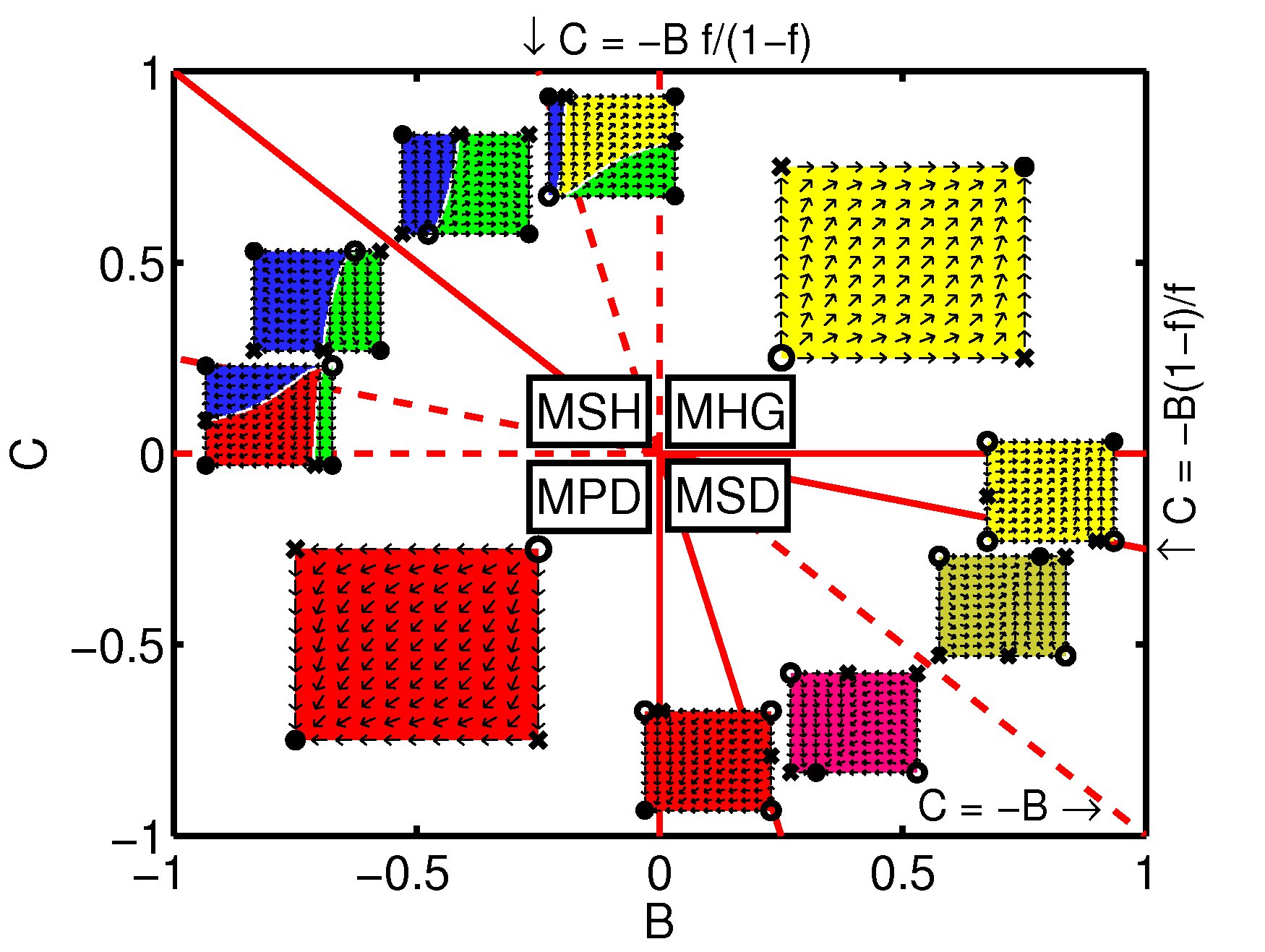}
\caption{(Color Online)
Illustration of the parameter-dependent types of outcomes as a function of the payoff-dependent parameters $B_a=b_a=B$ and $C_a=c_a=C$, if $f=0.8$ (i.e. 80\% of the entities belong to population 1) and if the entities have interactions with other entities, independently of the population they belong to. This corresponds to the multi-population case with interactions and self-interactions. Small arrows illustrate the vector field $(dp/dt,dq/dt)$ as a function of $p$ and $q$. Empty circles stand for unstable fix points (repelling neighboring trajetories), black circles represent stable fix points (attracting neighboring trajectories), and crosses represent saddle points (i.e. they are attractive in one direction and repulsive in the other). The basins of attraction of different stable fix points are represented in different shades of grey (colors) [red = (0,0), green = (1,0), blue = (0,1), yellow = (1,1), salmon = $(u,0)$, mustard = $(v,1)$, where $0<u,v<1$]. Solid red lines indicate the thresholds at which continuous phase transitions
take place, dashed lines indicate discontinous phase transitions. For multi-population prisoner's dilemmas (MPD), we have $B<0$ and $C<0$, for multi-population snowdrift games (MSD),  we have $B>0$ and $C<0$, for multi-population harmony games (MHG), we have $B>0$ and $C>0$, and for the multi-population stag hunt game (MSH), we have $B<0$ and $C>0$.}
\label{FIG3}
\end{figure*}

\subsection{Special Cases}

If there are no interactions between entities of different
populations, we have $B_a = 0 = C_a$. In that case, both populations
separately behave as expected in the one-population case (see Fig. 1 and Movie 1 \cite{Movies}). Instead, if there are interactions between both populations, but no self-interactions,
we have $b_a = 0 = c_a$. In that situation, we end up with
conventional bimatrix games (see Fig. 2 and Movie 2 \cite{Movies}). In the following, we will assume that everyone has interactions with entities of {\it all} populations with a
frequency that is proportional to the relative population sizes. For
simplicity, we will furthermore focus on the case where the
payoffs depends only on the state, but not the population of the interaction partner.
Then, we have $p_a = P_a = P$, $r_a = R_a = R$, $s_a = S_a = S$, and
$t_a = T_a = T$, i.e.
\begin{equation}
b_a = B_a = B = S-P \label{cas1}
\end{equation}
and
\begin{equation}
c_a=C_a = C = R-T  \label{cas2}
\end{equation}
(see Fig. 3 and Movie 3 \cite{Movies}). If the interaction rate between {\it different} populations is $\nu$ times the interaction rate within the {\it own} population, we have the
more general relationship $B_a = \nu b_a = \nu B$ and $C_a = \nu c_a
= \nu C$ (where the parameter $\nu >0$ allows us to tune the
interaction frequency between two populations---until now, we have
assumed $\nu=1$). In that case, we obtain
\begin{eqnarray}
F(p,q) = F_\nu(p,q) &=& B\underbrace{[f(1-p)  +\nu (1-f)q]}_{\ge 0} \nonumber \\
&+& C\underbrace{[fp + \nu (1 - f)(1-q) ]}_{\ge 0} \quad \label{gen1}
\end{eqnarray}
and
\begin{eqnarray}
G(p,q) =G_\nu(p,q) &=& B \underbrace{[(1-f)(1-q)  +\nu fp]}_{\ge 0} \nonumber \\
&+& C\underbrace{[(1-f)q + \nu f(1-p) ]}_{\ge 0}   \, . \quad \label{gen2}
\end{eqnarray}
Note that one can restrict the analysis of the two-population
game-dynamical equations to $f\ge 0.5$, as the transformations
$f\leftrightarrow (1-f)$ and $p\leftrightarrow q$ leave the two-population replicator
equations unchanged.

\section{Stationary Solutions, Eigenvalues, and Possible System Dynamics}\label{STAT}

In the two-dimensional space defined by the variables $p$ and $q$,
the qualitative properties of the vector field (which determines the
temporal changes $dp/dt$ and $dq/dt$) can be completely derived from
the stationary solutions and their stability properties, which are
given by their eigenvalues. These can be calculated
{\it analytically}, i.e. there are exact mathematical formulas for them.

\subsection{Basic Definitions}

For an interdisciplinary readership, we will shortly define some relevant terminology
here, while specialists may directly continue with subsection B.
A stationary solution $(p_l,q_l)$ is defined as a point with $dp/dt=0$ and $dq/dt=0$,
which implies
\begin{equation}
p_l(1-p_l)F(p_l,q_l) = 0 \quad \mbox{and}  \quad q_l(1-q_l)G(p_l,q_l)
= 0 \, .
\end{equation}
Besides calculating the stationary solutions, one may perform
a so-called {\it ``linear stability analysis''}, which allows one to
find out how a solution
\begin{equation}
(p(t),q(t)) = (p_l+\delta p_l(t),q_l+\delta q_l(t))
\end{equation}
in the vicinity of a stationary solution $(p_l,q_l)$
evolves in time. If the distance
\begin{equation}
d_l(t) = \sqrt{\delta p_l(t)^2 + \delta q_l(t)^2}
\end{equation}
goes to zero, which may be imagined as an
attraction towards the stationary solution, one speaks of a {\it
stable} stationary point or an asymptotically stable fix point
or an evolutionary equilibrium \cite{gamedyn3} (which
is a so-called {\it Nash equilibrium}).
Its {\it basin of attraction} is defined by the set of all initial conditions
$(p(0),q(0))$, for which the trajectories $(p(t),q(t))$ starting in
these points end up in the fix point under consideration as time
$t$ goes to infinity. (In Figs. 1--5 and Movies 1--3 \cite{Movies}, they are represented
by different background colors.)
\par
If the distance $d_l(t)$ grows rather than shrinks with time $t$,
one speaks of an {\it unstable} fix point. This may be imagined like
a {\it repulsion} from the stationary solution. If the growth or
shrinkage of the distance $d_l$ is a matter of the specific choice
of the initial conditions $p(0) = p_l +\delta p_l(0)$ and $q(0) =q_l
+ \delta q_l(t)$, the stationary point is called a {\it saddle
point}. A saddle point is attractive in one direction, but repulsive
in another one. In Figs. 1--5 and Movies 1--3 \cite{Movies}, the stationary points and
their respective stability properties (marked by circles and crosses) have
been determined {\it analytically}. They fit perfectly to the {\it numerically} calculated
vector fields, which represent $(dp/dt,dq/dt)$, i.e. the size and direction of changes in
the distribution $(p,q)$ of states with time.

\subsection{Calculation of the Stationary Solutions and their Eigenvalues}

We will now identify the stationary solutions $(p_l,q_l)$ satisfying
$dp/dt = 0$ and $dq/dt=0$ and their respective eigenvalues
$\lambda_l$ and $\mu_l$. Using the notation $p(t) = p_l + \delta
p_l(t)$ and $q(t) = q_l + \delta q_l(t)$, the eigenvalues follow
from the linearized equations
\begin{equation}
\frac{d}{dt} \left(
\begin{array}{c}
\delta p_l(t) \\
\delta q_l(t)
\end{array} \right)
= \left( \begin{array}{cc}
M_{11} & M_{12} \\
M_{21} & M_{22}
\end{array} \right)
\left(
\begin{array}{c}
\delta p_l(t) \\
\delta q_l(t)
\end{array} \right)
\label{eins}
\end{equation}
with
\begin{eqnarray}
M_{11} &=& (1-2p_l) F(p_l,q_l) + p_l(1-p_l) (c_1 - b_1) f \, , \nonumber \\
M_{12} &=& p_l(1-p_l)  (B_1-C_1)(1-f) \, , \nonumber \\
M_{21} &=& q_l(1-q_l) (B_2-C_2)f \, , \\
M_{22} &=& (1-2q_l) G(p_l,q_l) + q_l(1-q_l) (c_2 - b_2)(1- f) \, .
\nonumber
\end{eqnarray}
As the eigenvalue analysis of linear systems of differential
equations is a standard procedure \cite{gamedyn3}, we will not
explain it here in detail. We just note that the eigenvalues
$\lambda_l$ and $\mu_l$ of a stationary point $(p_l,q_l)$ are given
by the two solutions of the so-called {\it characteristic
polynomial}
\begin{equation}
(M_{11} - \lambda_l)(M_{22}-\mu_l) - M_{12}M_{21} = 0 \, .
\end{equation}
For the four stationary points $(p_l,q_l)$ with $l\in \{1,2,3,4\}$
discussed below, we have $p_l, q_l \in \{0,1\}$, which implies
$M_{12}M_{21}= 0$. Therefore, the first associated eigenvalue is just
\begin{equation}
\lambda_l = M_{11} = (1-2p_l) F(p_l,q_l) \, ,
\end{equation}
and the second associated eigenvalue is
\begin{equation}
\mu_l = M_{22} = (1-2q_l) G(p_l,q_l)\, .
\end{equation}
The following paragraph is again written for an interdisciplinary
readership, while specialists may skip it.
If both eigenvalues are negative, the corresponding stationary point
$(p_l,q_l)$ is a {\it stable fix point}, i.e.
``trajectories'' $(p(t),q(t))$ in the neighborhood (flow lines) are
{\it attracted} to it in the course of time $t$. If $\lambda_l$ and
$\mu_l$ are both positive, the stationary solution will be an {\it
unstable fix point}, and close-by trajectories will be {\it
repelled} from it. If one eigenvalue is negative and the other one
is positive, closeby trajectories are attracted in {\it one}
direction, while they are repelled in {\it another} direction. This
corresponds to a {\it saddle point}. If both eigenvalues are
positive, closeby trajectories are repelled from the stationary
solution. That situation is called an {\it unstable fix point}.
\par
Let us now turn to the discussion of the stationary solutions of Eqs.
(\ref{sat1}) and (\ref{sat2}) with the specifications (\ref{into1})
and (\ref{into2}):
\begin{itemize}
\item For the stationary solution $(p_1,q_1)=(0,0)$, we have the associated eigenvalues $\lambda_1 = b_1f + C_1(1-f)$ and $\mu_1 = b_2(1-f) + C_2f$.
\item The point $(p_2,q_2)=(1,1)$ is also a stationary solution and has the eigenvalues
$\lambda_2 = -[c_1f + B_1(1-f)]$ and $\mu_2 = -[c_2(1-f) + B_2 f]$.
\item The stationary solutions $(p_3,q_3) = (1,0)$ and $(p_4,q_4) = (0,1)$ exist as well. They have the eigenvalues $\lambda_3 = -[c_1 f + C_1(1-f) ]$,
$\mu_3 = b_2(1-f)  + B_2f$ and $\lambda_4 = b_1f  + B_1(1-f)$,
$\mu_4 = -[c_2 (1-f) + C_2f ]$.
\item If $0\le p_k \le 1$ and $0 \le q_k \le 1$ with
\begin{eqnarray}
p_5 &=& \frac{b_1f + C_1(1-f)}{(b_1 - c_1) f } \, , \\
p_6 &=& \frac{b_1f  + B_1(1-f)}{(b_1 - c_1) f} \, , \\
q_7 &=& \frac{b_2(1-f)  + B_2f}{(b_2 - c_2) (1-f)} \, , \\
q_8 &=& \frac{b_2(1-f)  + C_2f}{(b_2 - c_2) (1-f)}  \, ,
\end{eqnarray}
we additionally have stationary points $(p_5,q_5) = (p_5,0)$ with
$F(p_5,0)=0$, $(p_6,q_6) = (p_6,1)$ with $F(p_6,1)=0$, $(p_7,q_7) =
(1,q_7)$ with $G(1,q_7)=0$, and/or $(p_8,q_8) = (0,q_8)$ with
$G(0,q_8)=0$. These have the associated eigenvalues
\begin{equation}
\qquad \lambda_5 = p_5(1-p_5) (c_1 - b_1) f, \  \mu_5 = G(p_5,0) \, ,
\end{equation}
\begin{equation}
\qquad  \lambda_6 = p_6(1-p_6) (c_1 - b_1) f, \  \mu_6 = - G(p_6,1) \, ,
\end{equation}
\begin{equation}
\quad\mbox{ \ \,}  \lambda_7 = -F(1,q_7),  \ \mu_7 = q_7(1-q_7) (c_2 - b_2)(1- f) \, ,
\end{equation}
\begin{equation}
\qquad  \lambda_8 = F(0,q_8), \  \mu_8 = q_8(1-q_8) (c_2 - b_2)(1- f) \, .
\end{equation}
\item Inner stationary points $(p_9,q_9)$ with  $0< p_9 < 1$, $0<q_9<1$ can only exist, if $F(p_9,q_9) = 0 = G(p_9,q_9) $ can be satisfied.
\end{itemize}

\subsection{Special Case of Homogeneous Parameters}

Let us now focus on the case of homogeneous parameters given by $b_a=B_a = B$ and $c_a=C_a = C$.
In this case, the condition $F(p_9,q_9) = 0 = G(p_9,q_9)$ for an
inner point can only be fulfilled for $B+C=0$. If $B=-C$, one finds
a line
\begin{equation}
q(p) = \frac{1/2+f(p-1)}{1-f}
\end{equation}
of fix points, which are stable for $B>0$, but unstable for $B<0$.
Otherwise, fix points are only possible on the boundaries with either $p$ or
$q\in\{0,1\}$.
\par
Evaluating the conditions $0\le p_l \le 1$ and $0\le q_l\le 1$
reveals the following:
\begin{itemize}
\item The stationary point $(p_5,0)$ only exists for $C< 0< B$ and $f \ge |C|/(B+|C|)$ or for $B< 0< C$ and $f\ge C/(|B|+C)$.
\item $(p_6,1)$ is a stationary point for $C< 0< B$ and $f\ge B/(B+|C|)$ or for $B< 0< C$ and $f \ge |B| /(|B|+C)$.
\item The stationary point $(1,q_7)$ only exists for $C< 0< B$ and $f \le |C|/(B+|C|)$ or for $B< 0< C$ and $f\le C/(|B|+C)$.
\item $(0,q_8)$ is a stationary point for $C< 0< B$ and $f\le B/(B+|C|)$ or for $B< 0< C$ and $f \le |B|/(|B|+C)$.
\item If both, $B$ and $C$ are positive or negative at the same time, stationary points $(p_l,q_l)$ with $l\in\{5,\dots,8\}$ do not exist.
\end{itemize}

\section{Overview of Main Results}\label{MAIN}

For the special case with $b_a=B_a=B$ and $c_a=C_a=C$, our results
depend on the type of game, the sizes $|B|$ and $|C|$ of the
payoff-dependent model parameters, and the power $f$ of population 1
(e.g. its relative strength). They can be summarized as
follows:
For all values of the model parameters $B$, $C$, and $f$, all four corner points
(0,0), (1,0), (0,1), and (1,1) are stationary solutions. However,
if $B>0$ and $C>0$, the only asymptotically stable fix point is (1,1),
while for $B<0$ and $C<0$, the only stable fix point is (0,0).
In both cases, (1,0) and (0,1) are saddle points, and
stationary points $(p_l, q_l)$ with $l\in\{5, . . . , 8\}$ do not exist, as either the value of $p_l$ or
of $q_l$ lies outside the range $[0,1]$, thereby violating the normalization conditions.
\par
If $B<0$ and $C>0$, we have an equilibrium selection problem \cite{Preprint} and find:
\begin{itemize}
\item (0,1) and (1,0) are always asymptotically stable fix points.
\item (0,0) is a stable fix point for $|C|/|B| < \min[f/(1-f),(1-f)/f]$.
\item (1,1) is a stable fix point for $|C|/|B| > \max [f/(1-f),(1-f)/f]$.
\end{itemize}
If $B>0$ and $C<0$ we have:
\begin{itemize}
\item (1,0) and (0,1) are always unstable fix points.
\item (0,0) is a stable fix point for $|C|/|B| > \max[f/(1-f),(1-f)/f]$.
\item (1,1) is a stable fix point for $|C|/|B| < \min [f/(1-f),(1-f)/f]$.
\end{itemize}
Moreover, if $B$ and $C$ have different signs, stationary points $(p_l, q_l)$ with $l\in\{5, . . . , 8\}$ may occur:
\begin{itemize}
\item $(p_5,0)$ is a fix point for $|C|/(|B|+|C|) \le f$, i.e. $|C|/|B| \le f/(1-f)$.
\item $(p_6,1)$ is a fix point for $|B|/(|B|+|C|) \le f$, i.e. $|C|/|B| \ge (1-f)/f$.
\item $(1,q_7)$ is a fix point for $|C|/(B+|C|) \ge f$, i.e. $|C|/|B| \ge f/(1-f)$.
\item $(0,q_8)$ is a fix point for $|B|/(|B|+|C|) \ge f$, i.e. $|C|/|B| \le (1-f)/f$.
\end{itemize}

\subsection{Phase Transitions Between Different Types of System Dynamics} \label{PHASE}

\begin{figure}[htbp]
\includegraphics[width=9cm]{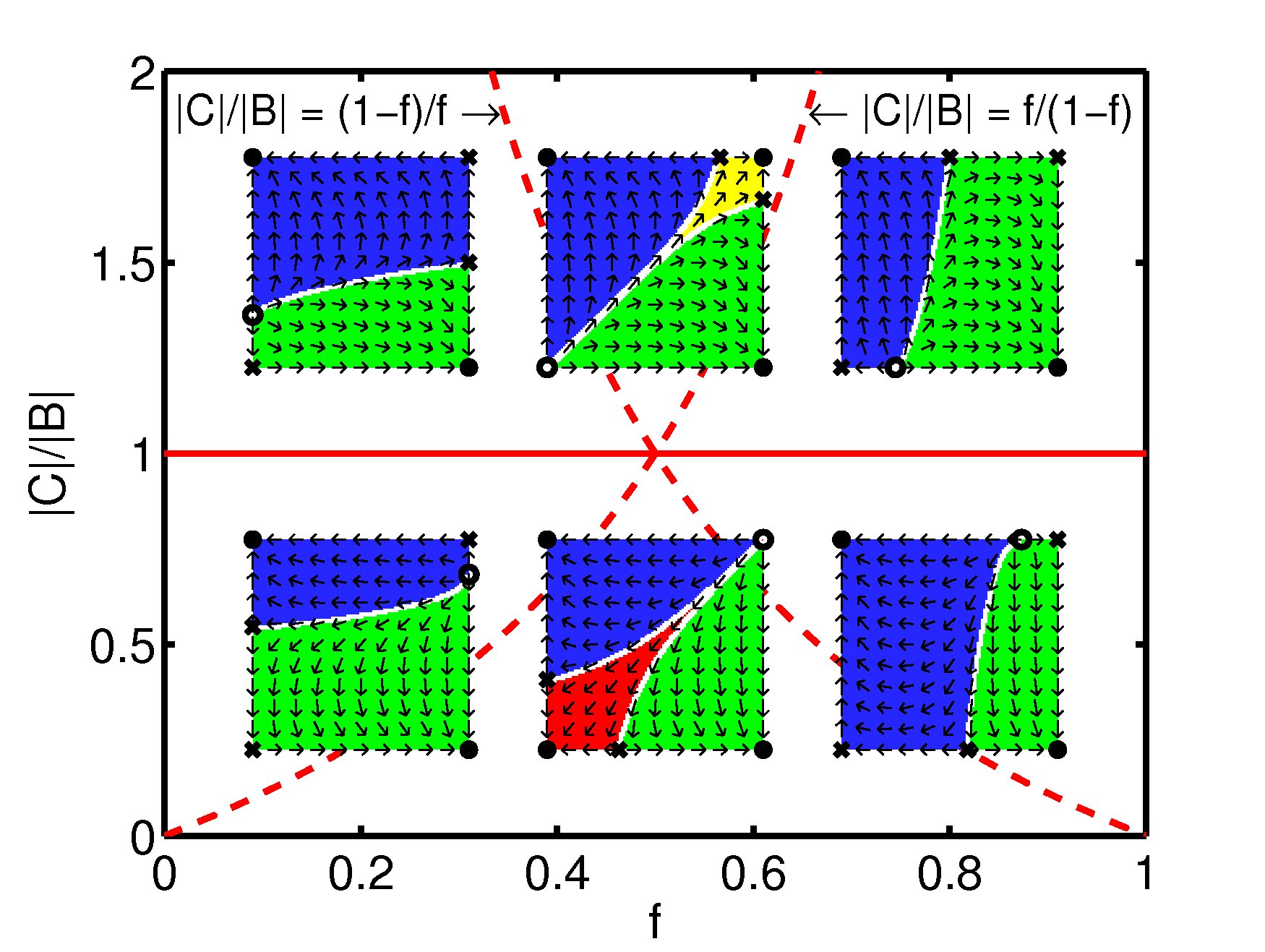}
\caption{(Color Online) Illustration of the parameter-dependent types of outcomes
in the multi-population stag-hunt game, if $|C|/|B|$ and/or $f$ are varied and
interaction between populations as well as self-interactions are considered.
The representation and grey shades (colors) are the same as in Fig. 3. Solid red lines indicate the thresholds at which continuous phase transitions take place, dashed lines indicate discontinous phase transitions.}
\label{FIG4}
\end{figure}

\begin{figure}[htbp]
\includegraphics[width=9cm]{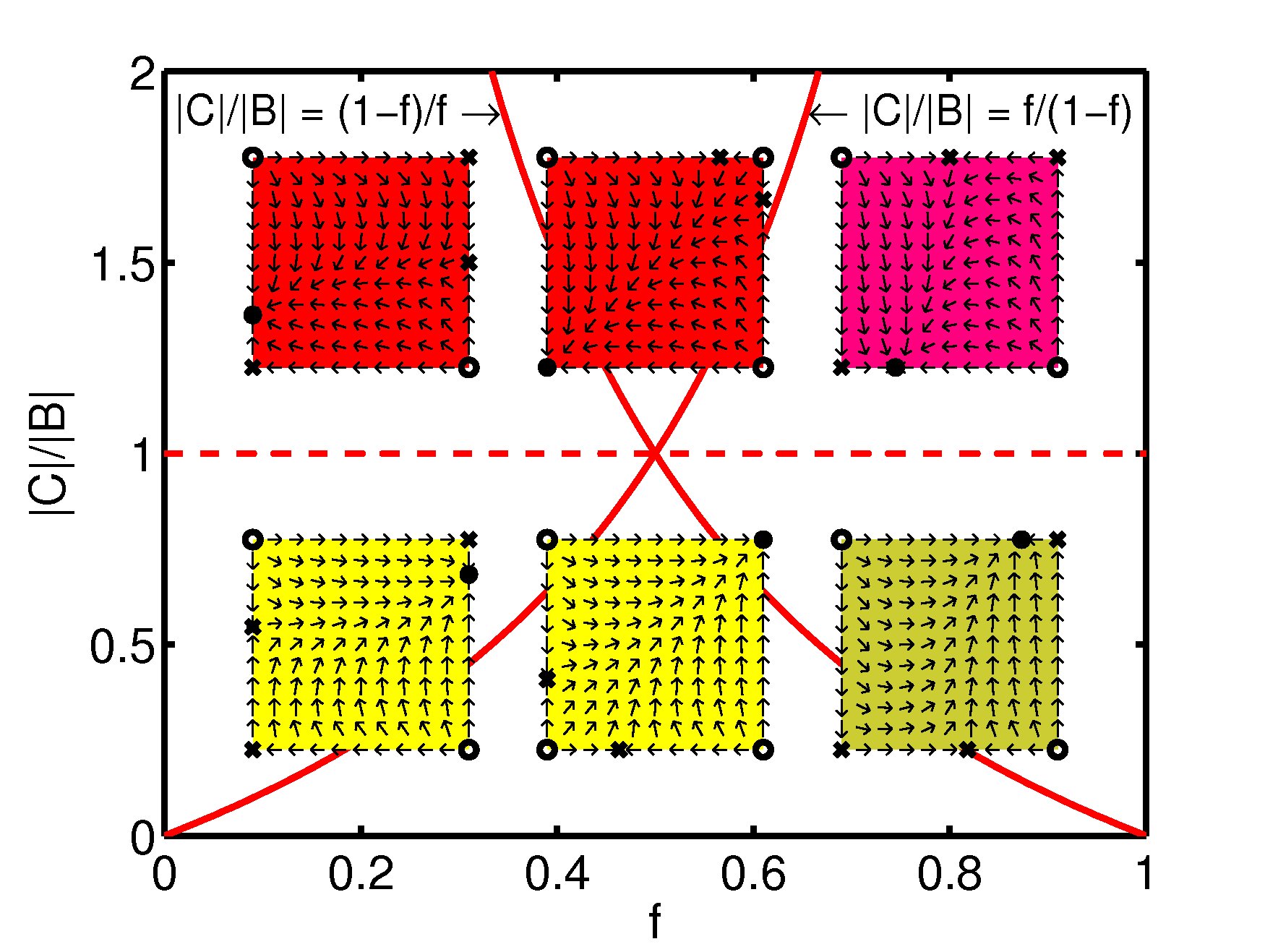}
\caption{(Color Online) Illustration of the parameter-dependent types of outcomes
in the multi-population snowdrift game, if $|C|/|B|$ and/or $f$ are varied and
interaction between populations as well as self-interactions are considered.
The representation and grey shades (colors) are the same as in Fig. 3. }
\label{FIG5}
\end{figure}

It is natural that a change in the parameters $B$, $C$, and $f$ causes changes in the system dynamics.
Normally, small parameter changes will imply smooth changes in the locations of fix points,
their eigenvalues, the vector fields, and basins of attraction.
However, when certain ``critical'' thresholds are crossed, new stable fix points may show up or disappear in remote
places of the parameter space, which defines a {\it discontinous (first-order) phase transition}.
If the locations of the stable fix points change {\it continuously} with a variation of the model parameters, while the related ``dislocation speed'' changes discontinously when crossing certain thresholds, we will talk of a {\it second-order phase transition}. In Figs. \ref{FIG1} to \ref{FIG5}, continuous transitions are indicated by solid lines, while discontinous transitions are represented by dashed lines.
\par
Analyzing the eigenvalues of the fix points (0,0), (1,0), (0,1), and (1,1), it is obvious that
our model of two populations with conflicting interactions shows phase transitions,
when $B$ or $C$ changes from positive to negative values or vice versa. The stationary
point (0,0) is stable for $B<0$ and $C<0$, (1,0) and (0,1) are stable for $B<0$ and $C>0$,
and (1,1) is stable for $B>0$ and $C>0$. This implies completely different types of
system dynamics, and the transitions between these cases are discontinous (corresponding
to first-order phase transitions). For $B>0$ and $C<0$, the stable fix point differs
from the corner points (0,0), (1,0), (0,1), and (1,1), but its location changes continuously,
as $B$ or $C$ crosses the zero line (corresponding to a second-order transition).
\par
It is striking that conflicting interactions between two populations lead to further transitions,
as $f$ or $|C|/|B|$ cross certain critical values:
Namely, as $|C|$ is increased from 0 to high values, apart from (0,0),
(0,1), (1,0), and (1,1), we find the following stationary points (given that $B$ and $C$ have different signs):
\begin{itemize}
\item $(p_5,0)$ and $(0,q_8)$, if $f\ge 1/2$ and $|C|/|B| \le (1-f)/f$ or if $f\le 1/2$ and  $|C|/|B| \le f/(1-f)$.
\item $(p_5,0)$ and $(p_6,1)$, if $f\ge 1/2$ and $(1-f)/f <|C|/|B| < f/(1-f)$, or $(1,q_7)$ and $(0,q_8)$ if  $f\le 1/2$ and $f/(1-f) <|C|/|B| < (1-f)/f$.
\item $(p_6,1)$ and $(1,q_7)$, if $f\ge 1/2$ and $|C|/|B| \ge f/(1-f)$ or if $f\le 1/2$ and  $|C|/|B| \ge (1-f)/f$.
\end{itemize}
For $B < 0 < C$, these fix points are unstable or saddle
points,  while they are stable or saddle points for $C < 0 < B$.
When the equality sign in the above inequalities applies, fix
points $(p_l,q_l)$ with $l\in\{5,6,7,8\}$ may become identical with
(0,0), (0,1), (1,0), or (1,1).
\par
Obviously, there are further transitions to a qualitatively different system behavior at the points $|C|/|B| =
(1-f)/f$ and $|C|/|B| =  f/(1-f)$ (see Figs. 3 to 5). These are continuous, if $B>0$ and $C<0$, but discontinuous for $B<0$ and $C>0$. Moreover, there is another transition, when $|C|$ crosses the
value of $|B|$, as the stability properties of pairs of fix points
are then {\it interchanged} (see Figs. 3--5 and Movie 3 \cite{Movies}).
If $B<0$ and $C>0$, this transition is of second order, as the stable fix points remain unchanged as the model parameters are varied (see Fig. 4). However, for $B>0$ and $C<0$, the transition is discontinuous (i.e. of first order), because the stable fix point turns into an unstable one and vice versa (see Fig. 5). That can be followed from
the fact that the dynamic system behavior and final outcome for the
case $|B|>|C|$ can be derived from the results for $|B| < |C|$. This is done
by applying the transformations $B\leftrightarrow -C$,
$p\leftrightarrow (1-p)$, and $q\leftrightarrow (1-q)$, which do not
change the game-dynamical equations
\begin{equation}
\frac{dp}{dt} = p(1-p) [Bf(1-p) + C f p + C(1-f)(1-q) + B(1-f)q]
\label{zus1}
\end{equation}
and
\begin{equation}
\frac{dq}{dt} = q(1-q) [B(1-f)(1-q) + C  (1-f) q + Cf(1-p) + B f p]  \, .
\label{zus2}
\end{equation}

\subsection{Classification and Interpretation of Different Types of System Dynamics}\label{CLASSI}

We have seen that the stability of the stationary points and the system dynamics change, when $B$ or $C$ cross the zero line. Therefore, it makes sense to distinguish four ``archetypical'' types of games. Note, however, that the two types with $BC <0$ can be subdivided into six subclasses each given by
\begin{itemize}
\item[(i)] $f/(1-f) < |C|/|B| < 1$,
\item[(ii)] $1 < |C|/|B| < (1-f)/f$,
\item[(iii)] $|C|/|B| < \min(f/(1-f),(1-f)/f)$,
\item[(iv)] $|C|/|B| > \max(f/(1-f),(1-f)/f)$,
\item[(v)] $(1-f)/f < |C|/|B| < 1$,
\item[(vi)] $1 < |C|/|B| < f/(1-f)$ (see Figs. 4+5).
\end{itemize}
That is, the system behavior for conflicting interactions (see Fig. 3) is clearly richer than for the one-population case \cite{Weibull,Preprint} or for two-population cases without interactions (see Fig. 1) or  without self-interactions (see Fig. 2). If $BC<0$, the system dynamics additionally depends on the values of $f$ and $|C|/|B|$. It may furthermore depend on the initial condition, if $B<0$ and $C>0$ (see Figs. 3+4).
\par
While our previous analysis has been formal and abstract, we will now discuss our results in the context of social systems for the sake of illustration. Then, the entities are {\it individuals}, and the states represent {\it behaviors}. Without loss of generality, we assume $R> P$ (determining the numbering and meaning of behaviors) and $f\ge 1/2$ (determining the numbering of populations such that the power of population 1 is the same or greater than the one of population 2). Moreover, we will use the following terminology: If two interacting individuals show the {\it same} behavior, we will talk about {\it ``coordinated behavior''}. The term {\it ``preferred behavior''} is used for the preferred {\it coordinated} behavior, i.e. the behavior which gives the higher payoff, when the interaction partner shows the {\it same} behavior. This payoff is represented by $R$, while the non-preferred coordinated behavior results in the payoff $P$. Furthermore, if a focal individual chooses its preferred behavior and the interaction partner chooses a different behavior, the first one receives the payoff $S$
and the second one the payoff $T$. In the so-called prisoner's dilemma, $R$ usually stands for
``reward'', $T$ for ``temptation'', $P$ for ``punishment'', and $S$ for ``sucker's payoff''. The payoff-dependent parameter $C=R-T$ may be interpreted as gain of coordinating on one's own preferred behavior (if greater than zero,
otherwise as loss). Moreover, $B=S-P$ may be interpreted as gain when
giving up coordinated, but non-preferred behavior.
\par
The conflict of interest between two populations is reflected by the fact that {\it ``cooperative behavior''}
is a matter of perspective: A behavior that appears cooperative to a focal individual is cooperative from the viewpoint of its interaction partner only, if belonging to the same population, otherwise it is non-cooperative from the interaction partner's viewpoint. In the model studied in this paper, population 1 prefers behavior 1, population 2 behavior 2. Moreover, behavior 1 corresponds to the cooperative behavior from the viewpoint of population 1, but to the {\it non-preferred} behavior of the interaction partner, i.e. it is {\it non-cooperative} from the point of view of population 2. Moreover, if two interacting individuals display the {\it same} behavior, their behavior is coordinated. Finally, we speak of a {\it ``behavioral norm''} or of {\it ``normative behavior'',} if all individuals (or the great majority) show the same (coordinated) behavior \cite{Epstein,Levin,Santos,Fent}, independently of their behavioral preferences and the (sub-)population they belong to. It should be stressed that this requires the individuals belonging to one of the populations to act against their own preferences. See Ref. \cite{JASSS} for the related social science literature.
\par
Within the context of the above definition, the four types of system dynamics distinguished above
are related to four types of games discussed in the following:
\begin{enumerate}
\item
For $T>R>P>S$ we have a {\it multi-population prisoner's dilemma
(MPD)}, which corresponds to the case $B<0$ and $C<0$. According to the results in
Sec. \ref{MAIN}, this is characterized by a breakdown of cooperation. Accordingly, individuals in both populations
will end up with their non-preferred behavior. This is even true, when the non-negative parameter $\nu$ in the generalized replicator equations (\ref{gen1}) and (\ref{gen2}) is different from 1.
\item
In contrast, for $R>T>S>P$ we have a {\it multi-population harmony
game (MHG)} with $B>0$ and $C>0$. In this case, all individuals end up with their
preferred behaviors, but the behavior of both populations is not coordinated.
Considering this coexistence of different behaviors, one could say that each population
forms its own ``subculture''.
\item
For $R>T > P>S$, which implies $B<0$ and $C>0$, we are confronted with
a {\it multi-population stag-hunt game (MSH)}. For most initial conditions, the system ends up
in the stationary states (1,0) or (0,1). In the first case, both populations coordinate themselves on the behavior
preferred by population 1, while in the second case, they coordinate themselves on the behavior preferred by
population 2. In both cases, all individuals end up with the same behavior. In other words, they establish a commonly shared behavior (a ``social norm''). However, there are also conditions under which {\it different} behaviors coexist, namely if (1,1) or (0,0) is a stable stationary point (see yellow and red basins of attraction in Fig. 4 and in the MSH section of Fig. 3). Under such conditions, norms are {\it not} self-enforcing, as a commonly shared behavior may not establish. This relevant case can occur only, if both populations have interactions {\it and} self-interactions. It should also be noted that norms have a similar function in social systems that forces have in physics. They guide human interactions in subtle ways, creating a self-organization of social order. See Refs. \cite{JASSS,HelEPJB} for a more detailed discussion of these issues.
\item
If $T>R>S>P$, corresponding to $B>0$ and $C<0$, we face a {\it multi-population
snowdrift game (MSD)}. In this case, it can happen that individuals in one of the populations
(the stronger one) do not coordinate among each other. While some of their individuals show a cooperative  behavior, the others are non-cooperative. We consider this fragmentation phenomenon as a simple description of social polarization or conflict.
\end{enumerate}
Note that, in the multi-population snowdrift game with $B>0$
and $C<0$, the stationary point $(p_5,0)$ exists for $f\ge |C|/(|B|
+ |C|)$, and the point $(p_6,1)$ for $f\ge |B|/(|B|+|C|)$. If $f\ge
1/2$ and $(1-f)/f <|C|/|B| < f/(1-f)$, $(p_5,0)$ is a stable fix
point for $|B|<|C|$, while $(p_6,1)$ is a stable fix
point for $|B| >|C|$, which implies a discontinuous transition at
the ``critical'' point $|B|=|C|$, when $|C|$ is continuously changed
from values smaller than $|B|$ to values greater than $|B|$ or vice
versa. This transition, where all individuals in the weaker population suddenly turn
from cooperative behavior from the perspective of the stronger population
to their own preferred behavior, may be considered to reflect a ``revolution''.
In the history of mankind, such revolutionary transitions have occured many times
\cite{Weidlich}.
\par
It turns out to be insightful to determine the {\it average}
fraction of cooperative individuals in {\it both} populations from
the perspective of the stronger population 1.
When $(p_5,0)$ is the stable stationary point, it can be determined as
the fraction of cooperative individuals in population 1 times the
relative size $f$ of population 1, plus the fraction $1-q_5=1$ of
non-cooperative individuals in population 2 (who are cooperative
from the point of view of population 1), weighted by its relative
size $(1-f)$:
\begin{eqnarray}
p_5 \!\cdot\! f + (1-q_5) \!\cdot\! (1-f)\!  &=& \! \frac{Bf+C(1-f)}{(B-C)f} \!\cdot\! f
+ 1 \!\cdot\! (1-f) \nonumber \\
&=& \!\frac{B}{B-C} = \frac{|B|}{|B|+|C|} \, .
\end{eqnarray}
Similarly, if $(p_6,1)$ is the stable
stationary point, the fraction of cooperative individuals from the
point of view of the stronger population 1 is given by
\begin{eqnarray}
p_6 \!\cdot\! f + (1-q_6) \!\cdot\! (1-f) \! & = & \! \frac{B}{(B-C)f} \!\cdot\! f + 0
\!\cdot\! (1-f) \nonumber \\
&=&\! \frac{B}{B-C} = \frac{|B|}{|B|+|C|} \, ,
\end{eqnarray}
as for $q_6=1$, everybody in population 2 behaves non-cooperatively
from the perspective of population 1. Surprisingly,  the average fraction of
cooperative individuals in both populations from the point of view
of the stronger population corresponds exactly
to the fraction $p_0 = |B|/(|B|+|C|)$ of cooperative individuals expected
in the one-population snowdrift game \cite{Preprint}. However, this comes with an enormous deviation
of the fraction $q$ of cooperative individuals in the {\it weaker}
population 2 from the expected value $p_0$ (as we either have $q=0$
or $q=1$), and also with {\it some} degree of deviation of $p$ from
$p_0$ in the stronger population 1. That is, although the stronger
population in the multi-population snowdrift game causes an
opposition of the weaker population and a polarization of
society \cite{Note}, the resulting distribution of behaviors in both
populations finally reaches a result, which fits the expectation of
the stronger population~1 (namely of having a fraction $p_0$ of
cooperative individuals from the point of view of population~1).
One could therefore say that the stronger population
controls the behavior of the weaker population.

\section{Summary and Outlook}\label{SUM}

In this paper, we have used multi-population replicator equations to describe
populations with conflicting interactions and different power. It turns out that the
system's behavior is much richer than in the one-population case or in the
two-population case without self-interactions. Nevertheless, it is useful to
distinguish four different types of games, characterized
by a qualitatively different system dynamics: The harmony game, the prisoner's
dilemma, the stag-hunt game and the snowdrift game. When applied to social systems,
the latter three describe social dilemma situations. However, in the presence of multiple
populations, we may not only have the dilemma that people may choose {\it not} to cooperate. Their
behaviors in different populations may also be {\it un}coordinated. Accordingly, the establishment
of cooperation is only {\it one} challenge in social systems, while the establishment of commonly
shared behaviors (``social norms'') is another one. Note that the evolution of social norms is
highly relevant for the evolution of language and culture \cite{Fortunato,Skyrms,Boyd}. According to our model, it is
expected to occur for multi-population stag hunt interactions. Interestingly, compared to the multi-population
games {\it without} self-interactions, we have found several new subclasses, depending on the power
$f$ of populations and the quotient $|C|/|B|$ of the
payoff-dependent parameters $B$ and $C$. The same is true for multi-population snowdrift games.
\par
Considering the simplicity of the model, the possible system behaviors are surprisingly rich.
Besides the occurrence of phase transitions when $B$ and $C$ change
their sign, we find additional transitions when $BC<0$ and the quotient $|C|/|B|$ crosses the values
of 1, $f/(1-f)$, or $(1-f)/f$. We expect an even larger variety of system behaviors, if the model
parameters are not chosen in a homogeneous way. For example, one could investigate cases in which
both populations play {\it different} games. Our model can also be extended to study
cases of migration and group selection. This will be demonstrated in forthcoming publications.
It will also be interesting to compare the behavior of test persons
in game-theoretical lab experiments \cite{Exp1,Exp2} with predictions of our model for interacting
individuals with conflicting interests. Depending on the specification of the interaction payoffs, it should be possible to find the following types of system behaviors: (1) The breakdown
of cooperation, (2) the coexistence of different behaviors (the establishment of ``subcultures''), (3) the evolution of commonly shared behaviors (``norms''), and (4) the occurence of social polarization.
In the latter case, one should also be able to find a ``revolutionary transition'' as $|B|/|C|$
crosses the value of 1. While there is empirical evidence that all these phenomena occur in real social systems, it will be interesting to test whether the above theory has also {\it predictive} power.

\subsection*{Author Contributions}
D.H. developed the concept and model of this study, did the analytical calculations and wrote the manuscript. A.J. prepared the figures and supplementary videos, and he performed the underlying computer simulations.

\subsection*{Acknowledgements}
The authors would like to thank for partial support by the ETH Competence Center ``Coping with Crises in Complex Socio-Economic Systems'' (CCSS) through ETH Research Grant CH1-01 08-2. They are grateful to Thomas Chadefaux, Ryan Murphy, Carlos Roca, Stefan Bechtold, Sergi Lozano, Heiko Rauhut, Wenjian Yu and further colleagues for valuable comments.
D.H. thanks Thomas Voss for his insightful seminar on social norms.


\begin{thebibliography}{99}
\bibitem{Note} Here, we understand ``polarization'' in the
sense that a population fragments into parts
with different behaviors.
\end{thebibliography}
\end{document}